\documentclass[manuscript, screen,authorversion,nonacm]{acmart}

\usepackage{todonotes}

\newtheorem{definition}{Definition}

\setcopyright{acmcopyright}
\copyrightyear{2020}
\acmYear{2020}
\acmDOI{XXX}

\acmConference[]{}{}{}
\acmBooktitle{}
\acmPrice{15.00}
\acmISBN{978-1-4503-9999-9/18/06}

\begin{document}

\title{Expressing Accountability Patterns using Structural Causal Models}

\author{Severin Kacianka}
\email{severin.kacianka@tum.de}
\orcid{}
\affiliation{%
  \institution{Technical University of Munich}
  \streetaddress{Boltzmannstr. 3}
  \city{Munich}
  \state{Germany}
  \postcode{85748}
}

\author{Amjad Ibrahim}
\email{amjad.ibrahim@tum.de}
\orcid{}
\affiliation{%
  \institution{Technical University of Munich}
  \streetaddress{Boltzmannstr. 3}
  \city{Munich}
  \state{Germany}
  \postcode{85748}
}

\author{Alexander Pretschner}
\email{alexander.pretschner@tum.de}
\orcid{}
\affiliation{%
  \institution{Technical University of Munich}
  \streetaddress{Boltzmannstr. 3}
  \city{Munich}
  \state{Germany}
  \postcode{85748}
}

\renewcommand{\shortauthors}{Kacianka, Ibrahim and Pretschner}
\renewcommand{\shorttitle}{Expressing Accountability Patterns using Structural Causal Models}

\begin{abstract}
While the exact definition and implementation of accountability depend on the
specific context, at its core accountability describes a mechanism that will
make decisions transparent and often provides means to sanction ``bad''
decisions.  As such, accountability is specifically relevant for Cyber-Physical
Systems, such as robots or drones, that embed themselves into a human society,
take decisions and might cause lasting harm. Without a notion of accountability,
such systems could behave with impunity and would not fit into society. 
Despite its relevance, there is currently no agreement on its meaning and, more
importantly, no way to express accountability properties for these systems. 
As a solution we propose to express the accountability properties of systems
using Structural Causal Models. They can be represented as human-readable
graphical models while also offering mathematical tools to analyze and reason
over them. 
Our central contribution is to show how Structural Causal Models can be used to
express and analyze the accountability properties of systems and that this
approach allows us to identify accountability patterns. These accountability
patterns can be catalogued and used to improve systems and their architectures.

\end{abstract}



\maketitle

\section{Introduction}
Accountability is a central pillar of human societies. Ever since John Locke and
Adam Smith, writers on political philosophy have tried to capture and refine
this slippery concept. Yet, details of definitions vary and there exists no
singular unified definition.  Following \cite{lindberg2013mapping}, who surveyed
the social science literature on accountability, the central idea of
accountability is that \emph{``when decision-making power is transferred from a
principal (e.g.  the citizens) to an agent (e.g. government), there must be a
mechanism in place for holding the agent accountable for their decisions and
tools for sanction''.} Accountability works by ensuring that agents might suffer
consequences for their behavior and will thus avoid ``bad'' behavior
\cite{mitchell1998power,hall2017psy_acc}. 

Causality, simply put, is about asking one question: Why?  It is at the core of
human understanding and reasoning, but has only recently been given thorough
mathematical foundation in the from of  \emph{Structural Causal Models}
(SCMs)~\cite{pearl2018book}. SCMs allow us to precisely describe and reason over
the causal structures of systems. They are uniquely suitable to capture
accountability properties of systems, because an agent can only be held
accountable for an effect that it caused; i.e., causality is a prerequisite for
accountability. Moreover, a recently published a survey of works in the social
sciences ~\cite{miller2018explanation} shows that Halpern and Pearl's model of
causality \cite{halpern2015} is a suitable formalization to capture the concept
of an ``explanation''. Explanations are tightly linked to accountability and
often seen as the ``account'' part of ``accountability''. 

As examples and use cases, we use Cyber-Physical Systems (CPS). CPS are systems
in which software controls hardware to affect the physical world
\cite{lee2008cyber}. Such systems usually have sensors to perceive the world,
employ actuators to influence their surroundings and use software to plan their
next action. CPS are open systems that are designed to work in unknown
environments and interact with any other system or object. This uncertainty
makes it impractical to specify all ``legal'' action a CPS can take and implies
that failures and unwanted events will happen. Similar to the transfer of
decision-making power from citizens to government, owners and operators of CPS
transfer some, and in some cases even all, their decision-making to the CPS.
Hence, we want similar \emph{accountability mechanisms} in place.  Such
accountability mechanisms help to avoid ``bad'' or unwanted behavior by ensuring
that its causes can be identified and there could be consequences.  Similar to
humans, the expectation of possible consequences will  motivate CPS and their
owners and operators to comply with existing laws and mores. Thus, accountable
CPS will improve society's trust in their behavior and increase their acceptance
in their respective social context. 

In this paper, we focus on the \textbf{problem} of expressing accountability
patterns and the related problem of what data to log.  Expressing accountability
is difficult, because  there are many differing definitions of accountability,
there is no agreement an what constitutes an ``accountable system'', and there
exists no common language to express accountability patterns. Our
\textbf{solution} uses the fact that accountability is build on the notion of
causality and leverages recent advances in the mathematical treatment of
causality to express accountability properties as SCMs. We use the structure of
SCMs to inform logging choices in the system design.  Our \textbf{contribution}
is (1)~to show how SCMs can be used to express accountability patterns, (2)~that
they can be used to analyze and improve the system design, especially regarding
the logging choices, and (3)~that this systematic treatment makes it possible to
identify accountability patterns and apply them to systems or system
architectures. 

\section{What is Accountability?}
\label{sec:accountability}

Across the sciences, there is no consensus on how accountability should be
defined. Its definition is always dependent on the context of a system or agent
and the expectations of its principals.  \cite{kacianka2017mapping} conducted a
systematic mapping study in computer science and found that there is no
agreement on a definition and that most research papers use a ``dictionary
definition'' of accountability, or give no definition at all. While
prescribing what accountability is (e.g., \cite{baldoni2018information} or
\cite{kusters2010accountability}) works for specific contexts, in this paper we
focus on how to best express its meaning in general.  In this section, we will
give a brief overview of accountability and related concepts such as causality,
responsibility and transparency. 

Following \cite{smc2018}, we see accountability as a relation between a system
(e.g., a CPS such as a drone or a robot; it might, in principle, also be human),
called an \textbf{agent} by \cite{lindberg2013mapping}, and a natural or
juridical person, called a \textbf{principal}, that the agent is accountable to.
The exact form of this relation depends of the definition of accountability
used.  Systems can be part of a ``stack'' and one technical system might monitor
and correct another system. However, ``things'', such as technical systems,
without legal standing cannot be principals. A principal serves as a link to
society and the legal system.  Having a principal that cannot be sued and be
held liable for actions and decisions would sever that link and violate a core
requirement for accountability\footnote{There are some ideas of giving legal
standing to machines (``e-person''), however this idea comes with its own set of
problems.  See for example \url{http://www.robotics-openletter.eu/}}.  This
argument is similar to the reasons why the executive in many countries cannot be
replaced with ``liquid democracy'' based voting systems. At some point a
decision must be justified and the decision maker be held accountable\footnote{
This point is eloquently made by J.S. Mill \cite{mill1861consideration}:
\emph{``Responsibility is null when nobody knows who is responsible; nor, even
when real, can it be divided without being weakened. To maintain it at its
highest, there must be one person who receives the whole praise of what is well
done, the whole blame of what is ill.''} }.  This view is further strengthened
by experiments  confirming  that \emph{``(...) people, not objects, are held
accountable for a negative outcome''} \cite{samland2016prescriptive}.

Accountability additionally requires the concept of actions that can or cannot
be taken by the agent and requires an account (or explanation) to be provided to
the principal. Actions are the linchpin that connects accountability to
causality. If an agents takes an action, it will cause effects. Not acting might
also cause effects. Principles judge these effects and might demand an account
from the agent. An agent that provides a meaningful account is accountable,
while an agent whose actions cannot be questioned or reviewed is not
accountable. An \textbf{accountability mechanism} is the means to hold the agent
accountable. 

\textbf{Causality} and accountability are tightly interlinked and have some
complex interactions. While causality is necessary for accountability
attributions, people will often use the term ``cause'' for both a causal
relationship and an accountability relationship \cite{samland2016prescriptive}.
The accountability hypothesis states that \emph{``(...) causal queries are
generally ambiguous. They might refer either to causal relations in the narrow
sense, or they might request assessment of moral accountability''}
\cite{samland2016prescriptive}. If humans understand ``cause'' to mean ``moral
accountability'', they will associate it, for example, with norm violations and
consider the subject's mental state.  Competing hypotheses suggest that norms
might alter the causal model (i.e., what causes what) or that they might alter
causal selection (i.e., what are possible causes). To illustrate the point:
\cite{samland2016prescriptive} conducted several experiments in which subjects
were presented with different vignettes and asked to attribute causality. In
their experiments they, for example, presented participants with a vignette in
which a plant lover employs two gardeners to look after his plants.  In this
scenario the gardeners could apply different fertilizers on the plants; one
would help the plants to flourish and one would make them perish. The gardeners
were explicitly told which fertilizer to use, i.e., a norm was established.
When the participants were asked to assess the causal effect of the fertilizer
on the plants, they, with one exception, did not attribute the death of the
plants to the fertilizer, but to the norm-violating gardener.  If the vignette
was altered to have one gardener intentionally use the wrong fertilizer he was
picked more frequently and if it was modified to say that the gardener
accidentally picked the wrong fertilizer, he was picked less frequently. 

Studies like this show how tricky it is to define the meaning of causal
relationship and accountability. They also highlight how important it is to
spell out the causal relationships in a system, to express expected causal
relationships and to compare them to actually observed behavior. In
Section~\ref{sec:modeling}, we show how SCMs are a tool to capture these
relationships. However, we want to caution that the actual modeling is highly
subjective and context sensitive and that reasoning over SCMs is always model
relative. 

Accountability is not limited to just answering questions, often called
\textbf{responsiveness}. The agent must be able to make a decision, because
\emph{``(a) puppet acting as an extension of someone else's will is not a
legitimate object of accountability''}\cite{lindberg2013mapping} . Here it is
also important to note that there is a lively debate if machines are even
capable of making decisions and if they can have agency (see for example
\cite{floridi2004morality,sayes2014actor,simon2015distributed}). In our view,
machines are objects, but can make decisions. While they might not, like a
human, suddenly decide to not do their task, they have enough degrees of freedom
and enough \textbf{autonomy}~\cite{clough2002metrics} to make accountability
necessary and meaningful.  For an automatic and deterministic system,
accountability makes little sense, as any blame will lie with the natural or
legal person who operated or constructed the system.  For such a system, fault
localization or debugging is enough. For accountability to make sense, systems
have to be autonomous agents that make decisions and explore counterfactual
scenarios without human intervention. 

It is important to be aware that it is very difficult to compare accountability
definitions and define ``levels of accountability''. The problem is that
definitions of accountability often disagree on properties and their meaning.
Even if we agree on properties, they are often nominal. For example, we should
be able to name an agent and a principal, but it is not easy to find a way to
compare two agents or find a relation between two agent-principal pairs such
that the excess of the agent in one pair makes up for the lack of the principal
in the other. For example, how should we decide if one person is a good
principal for a robot and compare them to another person's way of being a
principal of an autonomous car? SCMs and accountability patterns based on them
offer a solution to systematically compare accountability notions.  

Accountability will often be used in conjunction, and is sometimes confused,
with the term \textbf{responsibility}.  Following \cite{singh2018decision},
accountability is attributing responsibility in addition to the more traditional
definition of capturing the relation between principals and agents given by
\cite{lindberg2013mapping}. Seen this way, a responsibility assignment is the
output of an accountability mechanism.  Simultaneously,
\cite{lindberg2013mapping} points out that responsibility is a necessary
prerequisite for accountability.  You cannot be held accountable for an action
or effect for which you are not responsible. Here ``responsible'' is not seen as
an output, but is seen very close to the term ``caused''. The difference being
that you can also \emph{accidentally cause} an effect. As ``responsibility''
requires conscious thoughts and actions, you are not responsible for this effect
and thus also not accountable. A common example is to become unconscious while
driving a car and causing an accident. You caused the accident, but you are
neither responsible nor will you be held to account.  Reviewing the literature
in psychology, \cite{hall2017psy_acc}, find that while ``accountability'' and
``responsibility'' are used interchangeably by some researchers, they are
usually distinguished with accountability imposing the additional requirement of
an external audience.  In line with \cite{samland2016prescriptive},
\cite{hall2017psy_acc} also point to literature that sees responsibility as a
prerequisite for causality and causality as the core of accountability.
\cite{sytsma2012two}, on the other hand, hold the view that causality is a
normative concept and that \emph{``(c)ausal attributions are typically used to
indicate something more akin to who is responsible for a given outcome than who
caused the outcome in the descriptive sense of the term used by philosophers''}.
\cite{floridi2004morality} differentiate responsibility and accountability along
the line of intentionality. For them, a non-human agent can be held accountable,
but to be responsible for an action would require them to have actual
intentions.

In the recent works on algorithmic accountability  (e.g.,
\cite{diakopoulos2015algorithmic, singh2018accountability}),
\textbf{transparency} is also often linked to accountability, and sometimes even
seen ``as a type of accountability''. Here the idea is that if the inner
workings of machines and algorithms are open to the public, i.e., they are
transparent, we can attribute responsibility for failures. One problem, for
example, is that, in contrast to decisions of a committee of people,
transparency for source code or machine learning models is not necessarily
useful, because they are often next to impossible to understand. Another problem
for us is that transparency does not make the principle explicit. It roughly
follows the definition of ``felt accountability'' (see e.g.,
\cite{hall2017psy_acc}) that agents are accountable if they think that someone
might check their actions. While this is enough in some instances, we believe
that, especially for CPS, principals should be known.  \cite{ananny2018seeing}
give ten further limitation of transparency as a tool for accountability and
suggest these limitations might serve as requirements for algorithmic
accountability systems.   We believe that causal models offer a useful way to
fulfill those requirements and ensure the accountability of a system. Causal
models are especially good at facilitating explanations, which are often much
more useful than a fully transparent~system.

To summarize, causality is a relation between objects. Responsibility, for us,
requires some intention on part of the causing agent. Accountability is a
societal framework that involves autonomous agents and third parties.
Transparency, while close to accountability,  lacks the concept of a principal.
In the rest of this paper, we will show how SCMs can be used to express
accountability properties (Section~\ref{sec:formalizing}), to identify patterns
(Section~\ref{sec:modeling}) and to improve the system design
(Section~\ref{sec:examples}). 

\section{Expressing Accountability}
\label{sec:formalizing}

To illustrate how difficult it is to express the accountability properties and
find common features, we will use two entirely different examples. The first is
the story of Titus Manlius, as told by Livy (Book 8, chapter 7; translation
\cite{livy} and a summary \cite{pauw1991dramatic}). In 349 BC, Titus Manlius
Torquatus led a campaign against the Latins. To increase the military
discipline, he  ordered his troops to strictly obey orders, not to act on their
own and not to leave their post.  His son (and namesake) Titus Manlius, while on
orders to patrol, engaged an enemy commander who insulted, ridiculed and
challenged him to a duel.  After defeating the enemy heroically in the duel, he
rode back to his camp and proudly told his father of his exploits. Titus Manlius
summoned all his men and rebuked his son for abandoning his post and breaking
military discipline, summarizing the dilemma as \emph{``(...) ut aut rei
publicae mihi aut mei meorum obliviscendum sit (...)''} (roughly: ``(...) that I
must either forget the republic, or myself and mine (...)'').  With this he had
his son bound to a stake, beheaded him with an axe and had the body burned with
full military honors.

What does this bloody tale from ancient Rome now have to do with our modern
understanding of accountability? First of all, we have all major  components
present. Titus Manlius, the father, is the \emph{principal} who ensures a norm
is not violated. His son is the \emph{agent}, that (literally) gives an
\emph{account} of his actions. The execution of the son is then the ultimate
\emph{sanction} for violating the given norm. The \emph{accountability
mechanism} is a bit harder to identify. Its goal is to keep up the norm,
ensuring military discipline, and it is embedded in the military as its
reporting framework. 

As a second example, we look at the  fatal 2018 AD crash of an autonomously
driving Uber car~\cite{ubercrash}. In this accident, the car, equipped with
cameras, radars, a LIDAR and other sensors, was driving autonomously and hit a
pedestrian crossing the road. Crucially, the Uber developers disabled the car's
collision avoidance functionality, because it interfered with the computer
control. While the car had a safety driver, responsible to intervene in case of
system failures, the safety driver was focused on monitoring the system and did
not notice the pedestrian.  

Similar to the story of Titus Manlius above, we can identify components that can
be used for accountability. The car is the \emph{agent}, its video feeds and log
data are the \emph{account} and should provide an explanation. The immediate
\emph{sanction} taken by Uber was to stop their fleet of autonomous cars. Where
it gets tricky is the question of the \emph{principal}. Depending on how we want
to model the problem, the principal could be Uber, the safety driver, a
regulatory agency, a number of other entities, or multiple such entities at
once. Since the principal sets the norm, the norm depends heavily on who to
consider as principal. A regulatory body's goal might be to ensure road safety,
Uber might wish to avoid public relation disasters and so forth. Without a clear
way to express these intricacies, it is needlessly difficult to identify the
actual and intended principal.  Because a principal needs to know of its role to
conduct any oversight,  the knowledge of being a principal is a requirement for
accountability and the related logging choices. Expressing this fact clearly
allows regulatory bodies to spell out who they expect to look at accounts and
also gives manufactures, owner and operators the ability to show that they
comply with the accountability requirements and have the necessary logging
infrastructure.

\subsection{Structural Causal Models}

As both the story of Titus Manlius and the story of the Uber car have similar
features, such as a principal. We now want to look at definitions of
accountability and identify patterns that allow us to match them to these
features. This allows us to determine if a given pattern of accountability is
suitable for a given system or if another pattern might be a better fit.
We propose to use \emph{Structural Causal Models} (SCMs)
\cite{pearl2016causal}[p. 26f.] to capture the relevant accountability features
of systems and model the world they interact with.  

\begin{definition} \label{def:scm}
	\textbf{Structural Causal Model} \cite{pearl2016causal}\\
	A structural causal model $M$ is a tuple $M = \mathcal{(U,V,F)}$, where
	\begin{itemize}
	    \item $\mathcal{U}$ is a set of \emph{exogenous} variables,
		\item $\mathcal{V}$ is a set of \emph{endogenous} variables, 
		\item $\mathcal{F}$ associates with each variable $X \in
		\mathcal{V}$ a function that determines the value of $X$  given the 
		values of all other variables.
	\end{itemize}

\end{definition}

Every SCM is associated with a \emph{graphical causal model} called the
``graphical model'' or the ``graph''. While the graph does not include the
details of $\mathcal{F}$, its structure alone is enough to identify patterns and
causes. In an SCM, exogenous variables are external to the model, meaning that
we chose not to explain how they are caused.  They are the root nodes of the
\emph{causal graph} and are not descendant of any other variable.  Endogenous
variables, on the other hand, are descendants of at least one exogenous variable
and model components of our system and world for which we want to explain
causes. $\mathcal{F}$ describes the relationships between all those variables.
If we knew the value of every exogenous variable, we could use $\mathcal{F}$ to
determine the value of every endogenous variable. In a graphical model every
node represents an \emph{endogenous} variable and arrows represent functions
from $\mathcal{F}$ between those variables. SCM are derived from structural
equation models (SEMs) (e.g., \cite{lomax2004beginner}), but in contrast to
them, their relation have a direction. 

If we want to investigate the cause of Titus Manlius son's violation of the
norm, we might model the behavior of the enemy commander as an exogenous
variable, which we accept as given and external to the model. His son's reaction
to those insults are \emph{caused by} those insults and would thus be modeled as
an endogenous variable. Figure~\ref{fig:titus_cm} depicts the graphical model
for the story of Titus Manlius. In the SCM, $\mathcal{U}$ would contain
\emph{Insults (I)}, meaning that our model does not investigate why the enemy
commander chose to insult Titus Manlius' son. $\mathcal{V}$ would contain the
fact \emph{Titus Manlius' son reacted (TM)}, \emph{Engaged in a Duel (ED)} and
\emph{Broke Discipline (BD)}. For these endogenous variables, we can perform
causal reasoning and give explanations.  Note that $\mathcal{U}$  and
$\mathcal{V}$ should contain facts that can be observed and measured. If we
cannot measure some influence, but suspect it is there, we can model it as a
background~variable. 

In our example, $I$ is given by some exogenous input, meaning that our model
cannot explain it any further. For $TM$ the equation would be $TM =
\mathcal{F}_{TM}(I)$. Often this function will include an error term to account
for unknown background factors, called $U$ that might affect a variable, giving
us $TM = \mathcal{F}_{TM}(I,U_{TM})$; accordingly, $ED =
\mathcal{F}_{ED}(TM,U_{ED})$ and $BD = \mathcal{F}_{BD}(ED,U_{BD})$.  For
simplicity, we model the functions in $\mathcal{F}$ so that they just transmit
the value and take on either the value ``true'' or ``false''. In practice, the
functions in $\mathcal{F}$ can be any mathematical function and will often be
probability~distributions.

\subsection{Causal Reasoning}
\label{sec:reasoning}
\begin{figure}[h]
	\centering
	\includegraphics[width=0.25\textwidth]{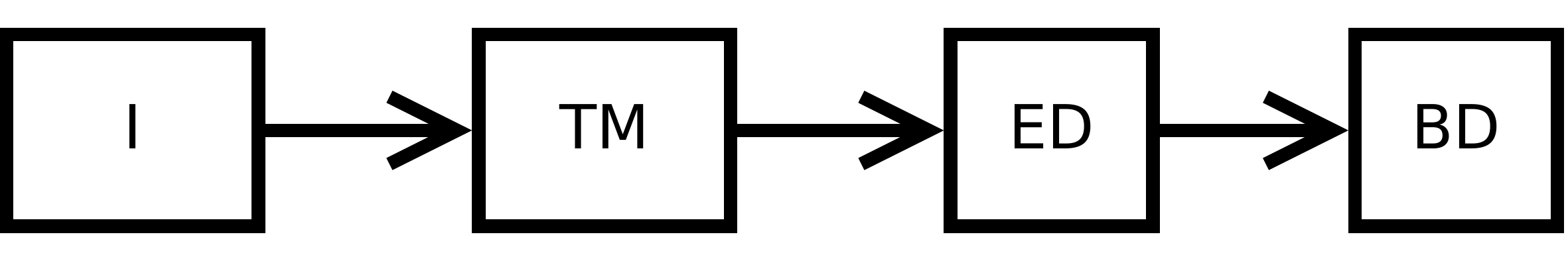}
	\caption{The story of Titus Manlius as graphical causal model. }
	\label{fig:titus_cm}
\end{figure}

Figure~\ref{fig:titus_cm} can now be used to answer causal
questions~\cite{pearl2018book}[p. 40]. The first level of questions are
questions of association, i.e., how one fact relates to another. For example, if
we know that the discipline was broken, did Titus Manlius' son engage in a duel?
To answer this question, we can inspect the graph and come up with the answer.
This task can of course also be automated \cite{tikka2018identifying} and the
graphs can be a lot more complex. 

\begin{figure}[h]
	\centering
	\includegraphics[width=0.25\textwidth]{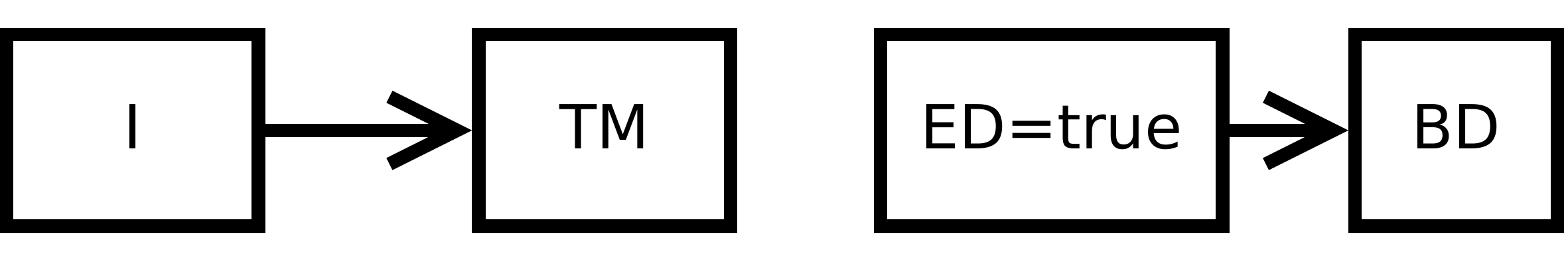}
	\caption{Simulating an intervention by fixing the value of an variable.}
	\label{fig:titus_cm3}
\end{figure}

Next, we might want to ask if the discipline is broken if anyone engaged in a
duel without being provoked. While questions of association can be answered with
standard logic, here we ask a hypothetical question about something we have no
data on. To analyze this, we first  remove all arrows leading into the node of
interest and, second, set this variable to the desired value. Then we can
analyze the resulting model with conventional logic.  The reason to do this is
that we want to analyze what would happen if we made a specific event happen.
Removing all arrows leading into the variable makes it independent from all
other influences. Forcing it to the desired value then ensures that we analyze
the world in which the event happens.  Figure~\ref{fig:titus_cm3} depicts this
model after the intervention. We disconnect $ED$ from all influences and then
set it to true. Analyzing the model, we can then see that this would also break
the discipline ($BD$). However, our model does not contain any information on
who might have engaged in the duel. If we wanted to answer this question, we
would need a new model containing additional variables.

\begin{figure}[h]
	\centering
	\includegraphics[width=0.25\textwidth]{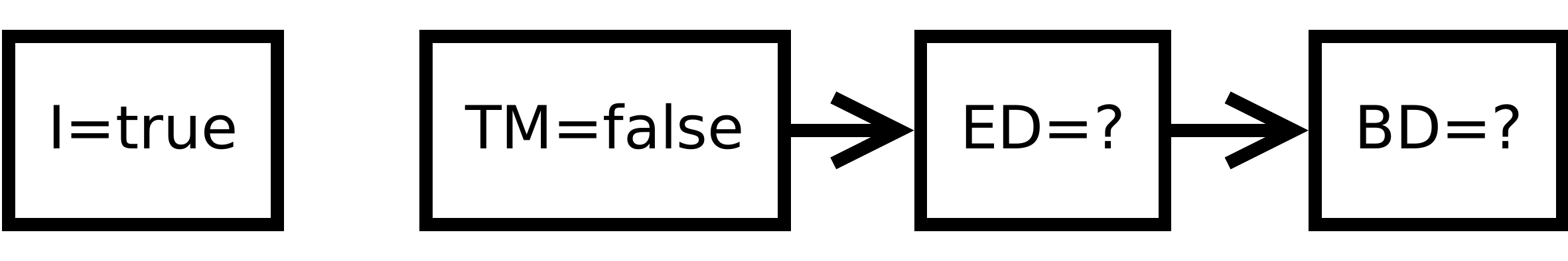}
	\caption{Counterfactual reasoning.  }
	\label{fig:titus_cm2}
\end{figure}

Finally, we can also use causal reasoning to ask counterfactual (``what if'')
questions. If we know that the discipline was broken, what would have happened
if Titus Manlius' son had not reacted to the insults? Would the discipline still
have been broken? Figure~\ref{fig:titus_cm2} depicts this model of the
counterfactual. We remove the link from $I$ to $TM$ and force its value to
``false'', i.e. to not be provoked. The variable $I$ keeps its original value.
The question then is, what would the value of $ED$ and $BD$ be? We could again
use automatic reasoning engines to deduce these values. 

Computing interventions and counterfactuals is what sets causal reasoning apart
from machine learning and other statistical approaches. If we were to use
machine learning, we would not have an explicit model of the world\footnote{Of
course, creating a causal model in the first place is a challenge.
\cite{crest,nfm} provide means to derive models for CPS from existing system
models. }. We would only gather one data point where the insult lead to a break
of discipline and a myriad of data points where there was no insult and also no
break in discipline. To understand Titus Manlius' causal contribution to the
effect, we need a model of the interactions and the ability to analyze it. SCMs
give us a way to express these models and then use well defined mathematical
operations to ask questions of association,  analyze interventions and
 compute counterfactuals. 

\subsection{Structures in SCMs}
\label{sec:structure}

To give a quick overview\footnote{For detailed introductions, see
\cite{pearl2018book} or \cite{pearl2014external}. }, a causal graph structure
such as a chain, $A \rightarrow B \rightarrow C$, means that $B$ is only
influenced by $A$, $C$ is only influenced by $B$ and $A$ is determined by
external forces not part of this model \cite{pearl2018book}[p. 129]. If we
inverted the chain, $A \leftarrow B \leftarrow C$, the causal meaning would
change, but the independence of the variables would remain the same, i.e., $A$
is still independent of $C$, provided we know $B$. Put another way, if there
were an arrow from $A$ to $C$, then changing $A$ would change $C$. However, if
there is no direct arrow and we keep the parents of $C$, i.e., $B$,
constant, no matter how we change $A$, $C$ will not be affected. A major
implication of this is that we can use data to test causal models. If we have
data that show that $C$ changes when we change $A$, despite $B$ not changing,
then we need to revise our model.
\cite{pearl2018book}[p.157] provides four
rules that apply to any causal graph: 

\begin{itshape} \begin{enumerate} \item In a chain junction, $A \rightarrow B
\rightarrow C$, controlling for $B$ prevents information about $A$ from getting
to $C$ or vice versa.

\item Likewise, in a fork or confounding junction, $A \leftarrow B \rightarrow
C$, controlling for $B$ prevents information about $A$ from getting to $C$ and
vice versa. 

\item In a collider, $A \rightarrow B \leftarrow C$, exactly the opposite rules
hold. The variables $A$ and $C$ start out independent, so that information about
$A$ tells you nothing about $C$. But if you control for $B$, then information
starts flowing through the ``pipe'', due to the explain-away effect. 

\item Controlling for descendants (or proxies) of a variable is like
``partially'' controlling for the variable itself. Controlling for a descendant
of a mediator partly closes the pipe; controlling for a descendant of a collider
partly opens the pipe.
\end{enumerate} 
\end{itshape}

In causal graphs, we often want to know if two nodes are independent, i.e.,
changes in one node will not affect another node. Two variables are  independent
if every path between them is \emph{blocked}. Formally:

\begin{definition}[d-seperation] \cite{pearl2016causal}[p. 46]

A path $p$ is blocked by a set of nodes $Z$ if and only if
\begin{enumerate}
\item $p$ contains a chain of nodes $A \rightarrow B \rightarrow C$, or a fork 
$A \leftarrow B \rightarrow C$, such that the middle node $B$ is in $Z$ (i.e.,
$B$ is conditioned on), or
\item $p$ contains a collider  $A \rightarrow B \leftarrow C$ such that the
collosion node $B$ is not in $Z$, and no descendant of $B$ is in $Z$. 
\end{enumerate}

If $Z$ blocks every path between two nodes $X$ and $Y$, then $X$ and $Y$ are
d-separated, conditional on $Z$, and thus are independent conditional on $Z$.  
\end{definition}

From do-calculus, we can also derive two graphical criteria. The front-door and
the back-door criterion allow us to analyze if a causal effect between two
variables is identifiable or not. Both of them can be checked automatically
\cite{tikka2018identifying}.  

\begin{definition}[The Back-Door Criterion]\cite{pearl2016causal}[p. 61] 
Given an ordered pair of variables $(X,Y)$ in a directed acyclic graph $G$, a
set of variables $Z$ satisfies the back-door criterion relative to $(X,Y)$ if no
node in $Z$ is a descendant of $X$, and $Z$ blocks every path between $X$ and
$Y$ that contains an arrow into $X$. 
\end{definition}

Intuitively, the back-door criterion ensures that (1) all spurious path between
$X$ and $Y$ are blocked, (2) all directed path from $X$ to $Y$ are not
perturbed, and (3) no new spurious paths are added.  

Building on the back-door criterion, the front-door criterion can applied in
structures not suitable for the back-door criterion.

\begin{definition}[The Front-Door Criterion]\cite{pearl2016causal}[p. 69]

A set of variables $Z$ is said to satisfy the front-door criterion relative to
an ordered pair of variables $(X,Y)$ if
\begin{enumerate}
\item $Z$ intercepts all directed paths from $X$ to $Y$.
\item There is no unblocked path from $X$ to $Z$.
\item All back-door paths from $Z$ to $Y$ are blocked by $X$.
\end{enumerate}

\end{definition}

Intuitively, the front-door criterion relies on the fact that one can identify
the effect of $X$ on $Z$ and the effect of $Z$ on $Y$ separately. Having
identified the separate effect, we can then calculate the effect from $X$ on
$Y$. 

\subsection{Accountability Patterns}
\label{sec:modeling}
Accountability becomes necessary  when a principal transfers power (for machines
often: delegates tasks) to an agent and that agent then causes some effect.  The
principal then has the right to understand how the agent used this power and why
it made a certain decision. We can now use SCMs to express definitions of
accountability and identify the underlying pattern.  The definitions given in
this section are abstractions and can usually not be used in real systems
directly. However, as they are condensed to their core, they are not obscured by
implementation details and can be used as patterns for real systems. If we
analyze the SCM of an actual system, we can look for these patterns, and if we
identify one, we can then use the knowledge about the pattern to improve the
actual system at hand.   Similarly to a design pattern in software engineering,
an accountability pattern is a reusable solution to a common problem, but must
be implemented anew in each system.  

Accountability patterns show us which nodes in a graph are actually relevant to
ensure a given notion of accountability. Using SCMs, we can then analyze how the
variables are connected and interact, and decide which variables need to be
logged. Without this clear understanding, we always run the risk of logging too
little data, or suffer from the burden of logging too much data, most of it
irrelevant. For example, if a path between two nodes is blocked, they are
independent and cannot influence each other. If one of those nodes is irrelevant
for accountability, we do not need to spend resources to log it. Looking at the
Uber example, the color of the Uber car was, presumably, irrelevant for the
accident and all accountability questions associated with it. So logging this
data would not help us clearing up any questions of accountability. 

In this paper, we use the definitions for  Lindberg- and RACI- accountability,
taken from \cite{smc2018}. As causality is at the core of accountability, we are
convinced that any other definition of accountability could be similarly
expressed as an SCM. Our two examples describe specific and very different
notions of accountability. We show that they can be expressed as a SCM, however,
we currently do not have a way to verify that the translation from the theory to
the model is correct. On the contrary, since accountability is a human and
social concept, the correctness of a model will always depend on a specific
context. The big advantage of causal models is that our assumptions about the
causal structure are explicit in the model and can be discussed and improved
upon by others. We consider SCMs ideally suited to express notions of
accountability and foster a discussion about their differences.

\subsubsection{Lindberg's Pattern}

Lindberg~\cite{lindberg2013mapping} surveyed the social science literature  and
provided the following synthesized definition of accountability:
\begin{itshape}
\begin{enumerate}
\item An agent or institution who is to give an account (A for agent);
\item An area, responsibilities, or domain subject to accountability (D for
domain);
\item An agent or institution to whom A is to give account (P for principal);
\item The right of P to require A to inform and explain/justify decisions with 
regard to D; and 
\item The right of P to sanction A if A fails to inform and/or explain/justify 
decisions with regard to D.
\end{enumerate}
\end{itshape}

\begin{figure}[h!]
	\centering
	\includegraphics[width=0.35\textwidth]{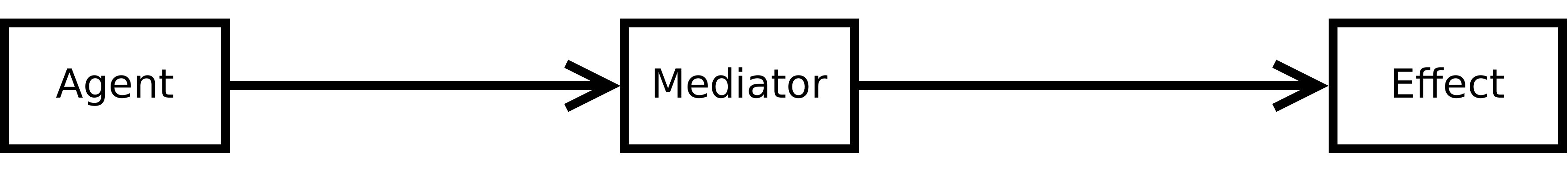}
	\caption{The causal model for the Lindberg accountability pattern; the
	principal is not part of the pattern. }
	\label{fig:lindberg_cm}
\end{figure}

Lindberg's definition assumes that an agent in an organization will cause some
effect, usually indirectly through mediation\footnote{ A mediator is a variable
between a cause and an effect. It might amplify or modify the initial input.
\cite{baron1986moderator} define mediation as a \emph{``generative mechanism
through which (...) [an] independent variable is able to influence the dependent
variable of interest''}. } (see Figure~\ref{fig:lindberg_cm}). If our system, or
a part of our system, has such a structure, regardless of the actual functions,
we can apply Lindberg's definitions to set up a framework that allows some
principal to get more information from the agent, i.e., to improve the cause
model. The principal is not part of the model, but will use the output of a
system following that model to sanction an agent.  In the same vein,  term
\emph{felt accountability}~\cite{hall2017psy_acc} is used in psychology to mean
that actors think there is a possibility that their actions will be evaluated by
a third party. 

\subsubsection{RACI Pattern}
In contrast to social sciences,  organizational sciences often apply a
\emph{Responsible-Accountable-Consult-Inform (RACI)} framework
\cite{smith2005role} to visualize the roles of people in an organization. The
elements of the framework are described as follows (adapted from
\cite{smith2005role}):
\begin{itemize}
\item \textbf{R}esponsible: The individual who completes a task. Responsibility
can be shared.
\item \textbf{A}ccountable: The person who answers for an action or decision.
There can be only one such person.
\item \textbf{C}onsult: Persons who are consulted prior to a decision. 
Communication must be bidirectional. 
\item \textbf{I}nform: Persons who are informed after a decision or action is
taken. This is unidirectional communication.  

\end{itemize}

\begin{figure}[h!]
	\centering
	\includegraphics[width=0.5\textwidth]{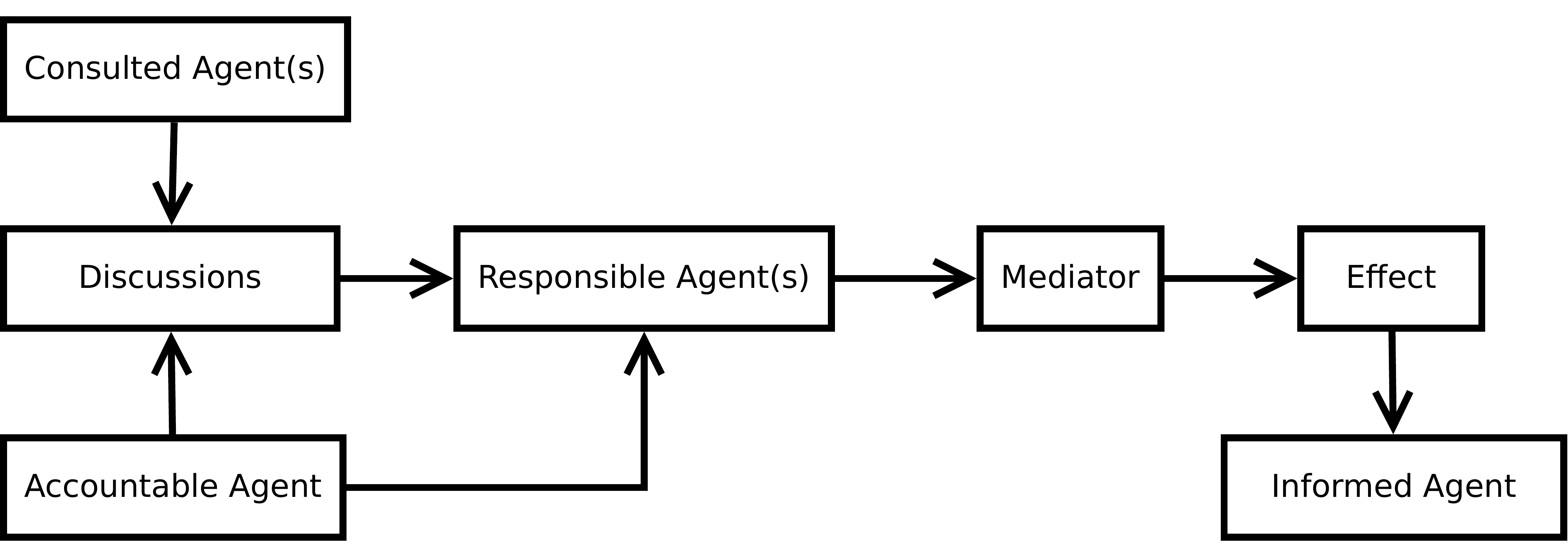}
	\caption{The RACI accountability pattern. }
	\label{fig:raci_cm}
\end{figure}

Figure~\ref{fig:raci_cm} shows the accountability pattern based on the RACI
framework. In contrast to Lindberg above, RACI includes variables that provide
additional information. So, instead of just needing data on the cause, the
mediator, and the effect, we additionally need information on the
\emph{accountable agent}, any \emph{consulted agent(s)}, the content of the
consultation, called \emph{discussion} here, and the \emph{informed agent(s)}.
While this pattern has much higher logging and reporting requirements, it also
allows us to model the delegation of decisions and common workflows in
organizations, like meetings, that cannot be captured in the Lindberg model. 

\section{Using Accountability Patterns}
\label{sec:examples}

\begin{figure}[h]
	\centering
	\includegraphics[width=0.25\textwidth]{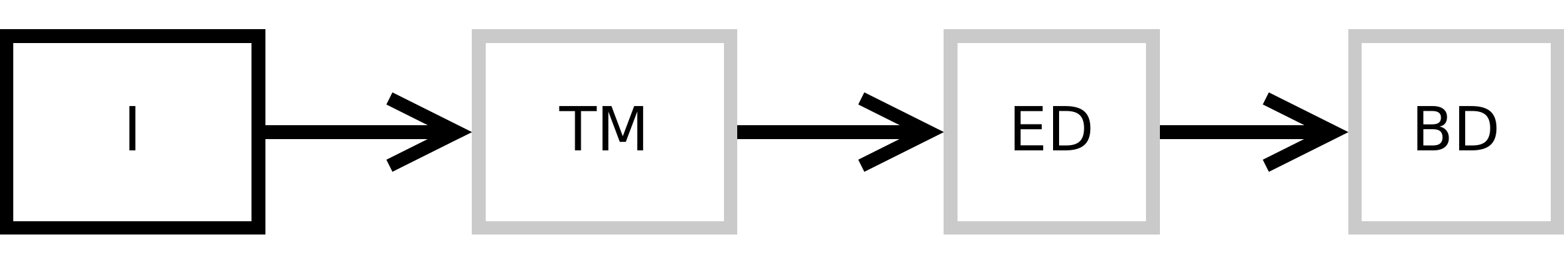}
	\caption{The story of Titus Manlius as causal model, with Lindberg
	accountability pattern highlighted in grey. }
	\label{fig:titus_cm_ex}
\end{figure}

We now show a possible causal model for the story of Titus Manlius and the Uber
example. If we find that they comply with a know accountability pattern, we can
be certain that the causal effect of the agent can be identified and know which
events we need to log.  
%
%
%
Figure~\ref{fig:titus_cm_ex} depicts
the familiar model of Titus Manlius' story. Here we highlight the structure of
the Lindberg accountability pattern in grey.  A mapping to the RACI pattern would
be far less convincing, mainly because Titus Manlius' son did not consult anyone
or delegated the task and we could not map the nodes required by the RACI
pattern. He was not an organization, which is why a model for organizational
accountability is no natural fit.  Knowing that in the Linderberg pattern, we
need to look at the behavior of the agent, we can, as shown in the examples on
causal reasoning in Section~\ref{sec:reasoning}, ask the counterfactual question
``What would have happened if  Titus Manlius' son had not engaged in a duel?''.
As in this case the military discipline would not have been broken, holding
Titus Manlius's son to account is the correct action.

\begin{figure}[t]
	\centering
	\includegraphics[width=0.45\textwidth]{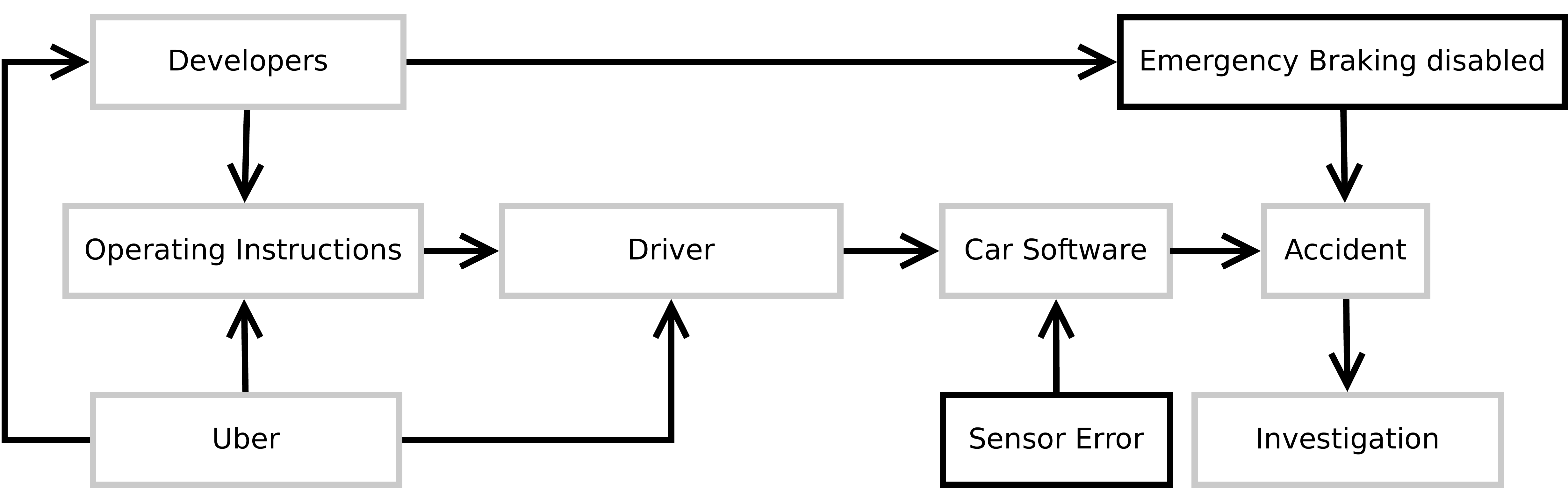}
	\caption{A causal model for the Uber accident with the RACI pattern
	highlighted in 	grey. }
	\label{fig:uber}
\end{figure}

Figure~\ref{fig:uber} models the Uber accident. In it, we depict that
Uber hired and trained a safety driver for its cars. The safety driver is aided
by operating manuals and procedures set by Uber's experts. Only with these
prerequisites the driver is allowed to let the car drive autonomously. Still,
this led to an accident that was reported to the police. In contrast to the
story of Titus Manlius, this model fits very nicely to the RACI patten. With Uber
being an organization, this is of course very natural. 

Similar to above, after an accident we can ask different counterfactual
questions. To find the responsible agent, we could ask ``Would the accident have
happened if the driver had not started the car?'', and, looking at the model, we
can see that the driver is on one direct path to the accident, so changing
\emph{Driver} would affect \emph{Accident}. However, there is also a path from
\emph{Developers} via \emph{Emergency Braking disabled} to \emph{Accident}. In
such a case, we say that the effects of the two nodes are \emph{confounded},
i.e., without additional data it is impossible to say what influence each node
has. The accountable entity, on the other hand, is unambiguous: If we change the
value of \emph{Uber}, all child nodes would change, clearly preventing the
accident.  If we chose the Linderberg accountability pattern for this graph, the
nodes \emph{Driver} would be the agent, \emph{Car Software} the mediator and
\emph{Accident} the effect.  In this case, we could not make an unambiguous
attribution of accountability and would need to either change the accountability
pattern or change the system to fit the desired pattern.

Of course, many other models are possible for these examples and a plethora of
competing definitions of accountability can be applied. However, our goal is not
to find the ``one true model''; our goal here is to show that causal models are
a useful tool to express and identify accountability pattern and analyze if a
systems fulfils them.  To stick with our examples, a regulatory body could
publish the RACI pattern and then check if Uber confirmed to it. If, on the
other hand, Uber said they used the Lindberg pattern, this could be criticized
and another pattern be proposed.  Without SCMs to express these fine points, a
discussion is next to impossible, and, most importantly, will lose the technical
details. This example also illustrates that SCMs without $\mathcal{F}$ can
answer some queries, but they will often not be able to give clear answers.
Defining $\mathcal{F}$ is one of the major open challenges when working with
SCMs. 

\subsection{Choosing an Accountability Pattern}
In the previous section, we showed that the Lindberg pattern would not be a good
fit for the Uber example. In this section, we will look at the SCM for a part of
a system and will show how we can use the strucures in the SCM
(Section~\ref{sec:structure}) to check if an accountability pattern fits the
model. As in the Uber example, for a pattern to be applicable, it is important
that its causal effect is identifiable and not confounded. As an example we use
another CPS and discuss the development of an accountability mechanism for an
Unmanned Aerial Vehicle (UAV). The goal is to find out who is accountable if the
UAV crashes in bad weather for which the pilot does not have the necessary
flight permissions.

\begin{figure}[h]
	\centering
	\includegraphics[width=0.45\textwidth]{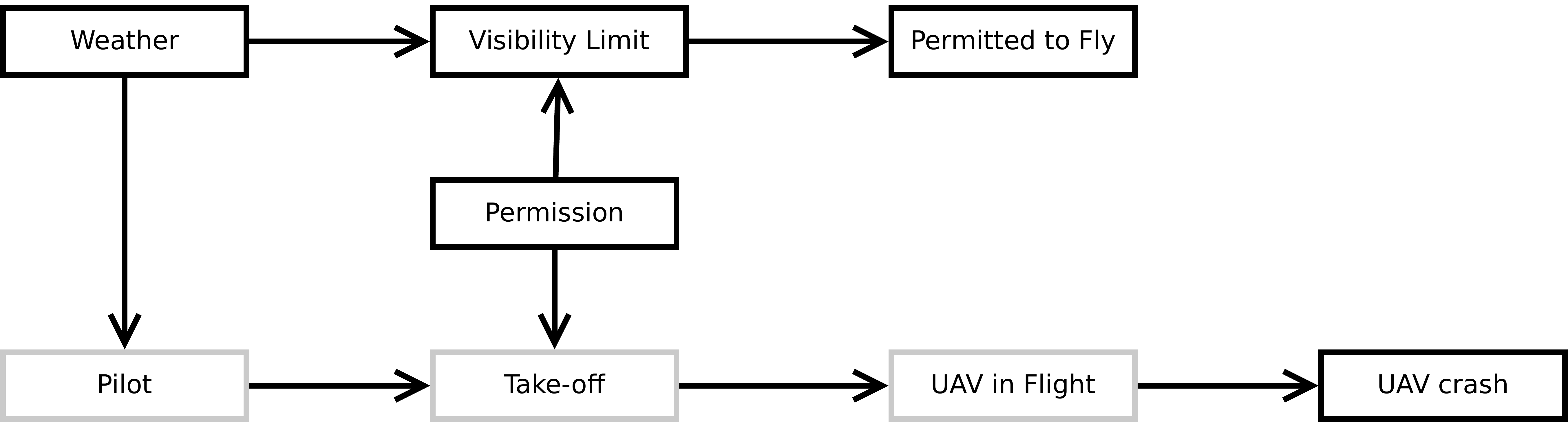}
	\caption{Causal Model with the accountability model highlighted in grey. }
	\label{fig:practical}
\end{figure}

Figure~\ref{fig:practical}\footnote{This model structure is similar to the
second structure given by \cite{pearl2018book}[p.159]. } depicts one possible
causal model for our UAV; the Lindberg pattern is highlighted in grey. It
captures four ``pre-flight'' variables, \emph{weather}, the legal
\emph{visibility limit}, the type of \emph{permission} obtained (e.g., allowed
to fly in bad weather) and resulting from those, \emph{permitted to fly}, that
signals whether a flight was legally allowed or not.  Additionally, it models
the pilot's decision, the fact that the UAV took off, that it is in flight and
whether it crashed. 

Under the assumption that this model reflects reality correctly, we can now use
the back-door criterion to decide what data to log and what data to use for
analysis. We want to make sure that the effect of the pilot on the crash is
identifiable and not confounded by the other factors.  To ensure that, we need
to make sure that every noncausal path is blocked, while not perturbing any
causal path~\cite{pearl2018book}[p.  158].  A back-door path between two nodes
$X$ and $Y$ is ``\emph{any path from $X$ to $Y$ that starts with an arrow
pointing into $X$''} \cite{pearl2018book}[p.  158].  The effect of $X$ on $Y$
will be de-confounded if every back-door path is blocked. In
Figure~\ref{fig:practical} we have one such path, \emph{Pilot} $\leftarrow$
\emph{Weather} $\rightarrow$ \emph{Visibility Limit} $\leftarrow$
\emph{Permission} $\rightarrow$ \emph{Take-off} $\rightarrow$ \emph{UAV in
flight} $\rightarrow$ \emph{UAV crash}. Looking at rule~3 above, we can see that
the \emph{Visibility Limit} is a collider, meaning we do not need to control for
(log) anything except \emph{Pilot}, \emph{Take-off}, \emph{UAV in flight} and
\emph{UAV crash}.  This finding is in line with our intuition: while the weather
might have an influence on the pilot, it also cannot make the UAV take off and
crash. The permission status might influence the take-off decision, but in
itself it cannot prevent a UAV from taking off. As child nodes of \emph{Weather}
and  \emph{Permission}, \emph{Visibility Limit} and the resulting fact that a
flight is permitted or not have even less influence on the observed behavior of
the UAV. Note that other cause for a crash, such as an engine failure, are
impossible in this model. To consider them, we would need to extend the model
\cite{nfm}.  

\begin{figure}[h]
	\centering
	\includegraphics[width=0.45\textwidth]{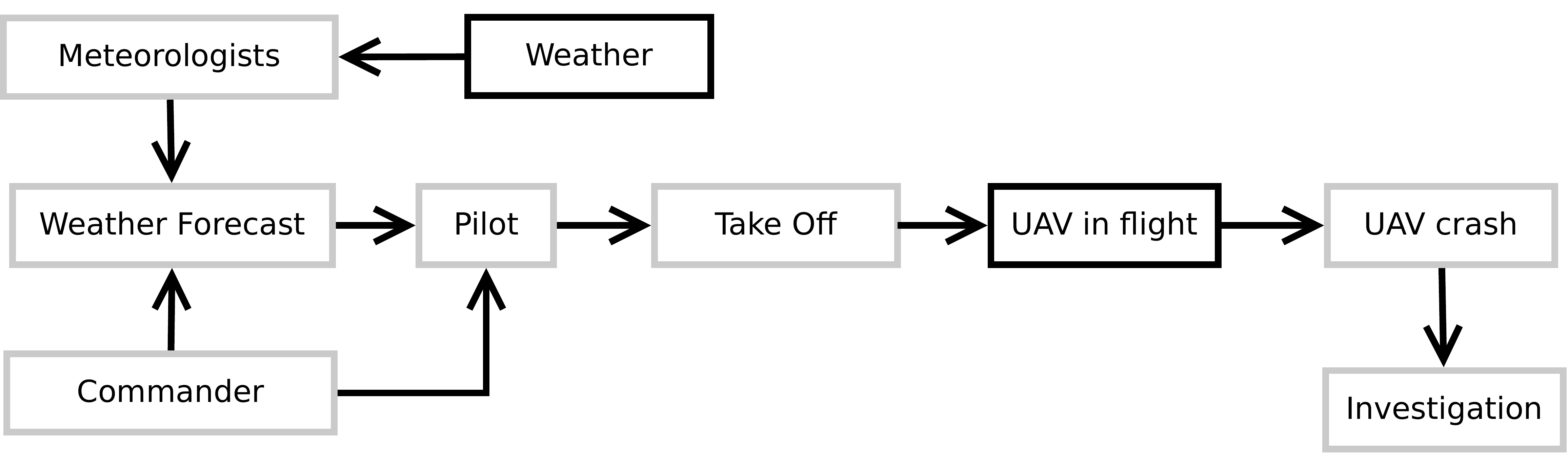}
	\caption{Using RACI to model the bad weather example. }
	\label{fig:bad_weather_raci}
\end{figure}

If we do not want to model hobby pilots who can start their UAVs, possible
disregarding laws and regulations, where- and whenever they want, and want to
capture a more formal military setup, we might design our overall system to
follow the RACI accountability pattern. Figure~\ref{fig:bad_weather_raci}
depicts such a setup. In such a RACI setting, pilots have less freedom of
choice.  They may only start the UAV if they receive an order to do so. This
order is given by a \emph{commander}, after consulting with
\emph{meteorologists} who observe the \emph{weather} and prepare a \emph{weather
forecast}. If the UAV is still lost due to bad weather, the pilot will still be
responsible, but accountability will, following the RACI definition, lie with
the commander giving the take-off order.  As in the previous example, we have a
collider in \emph{weather forecast}, that ensures that the \emph{commander} and
the \emph{meteorologist} are independent.

\subsection{Accountability Patterns and System Design}
\label{sec:system_design}

\begin{figure}[h]
	\centering
	\includegraphics[width=0.45\textwidth]{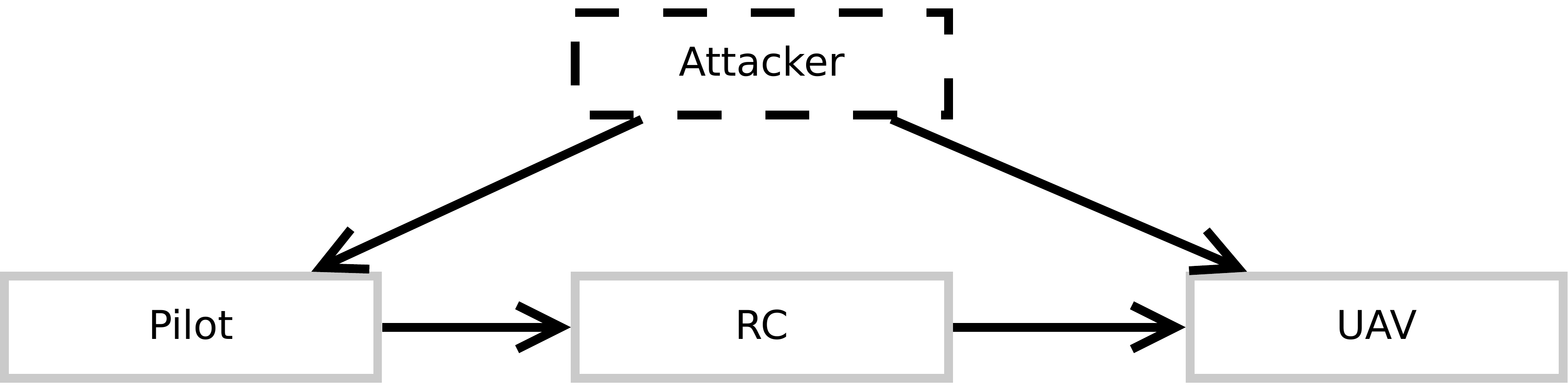}
	\caption{The unobservable attacker will confound the effect of the pilot on
	the UAV.}
	\label{fig:practical2}
\end{figure}

Here, we will show  how accountability patterns can help us when designing a
system. As an example, suppose we are tasked with attributing accountability in
the case where a UAV is controlled by a pilot, but both the pilot and the UAV
might be influenced by some attacker. To complicate matters, the attacker is an
unobservable confounding variable. That means that it will affect two observed
variables, but can itself not be observed. As such it might confound the effect
of the pilot on the UAV, which is a problem for the Linderberg accountability
pattern. Figure~\ref{fig:practical2} depicts the corresponding causal model.  In
contrast to the previous examples, there is the open path, \emph{Pilot}
$\leftarrow$ \emph{Attacker} $\rightarrow$ \emph{UAV}. Looking at rule~2, we
could close that path by controlling for the \emph{Attacker}. One option to do
this would be to install an intrusion detection system (IDS) and make the
\emph{IDS} a proxy for the \emph{Attacker}. In an ideal world, a \emph{proxy
variable} will behave like the real variable, but will in reality often induce
some uncertainty (e.g, because it might miss some attacks). 

\begin{figure}[h]
	\centering
	\includegraphics[width=0.45\textwidth]{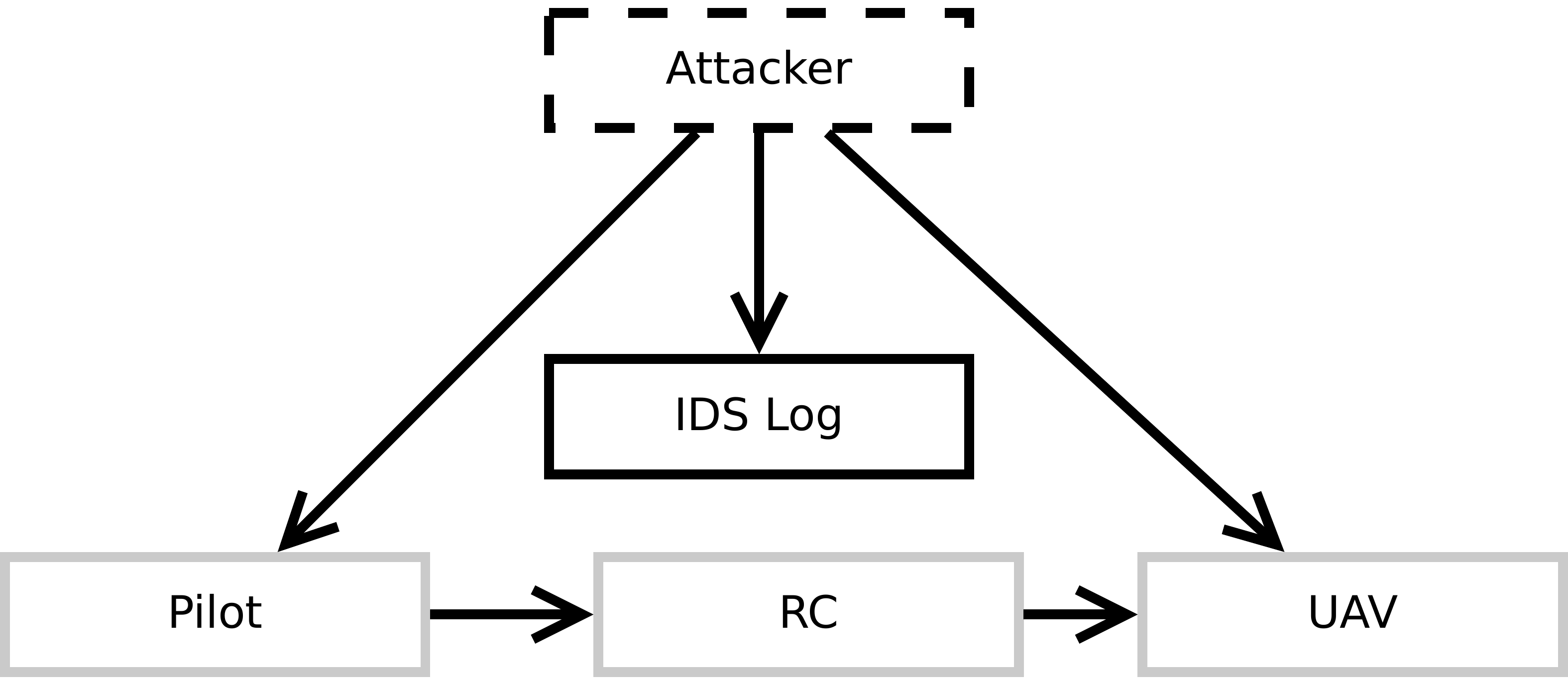}
	\caption{Using an IDS to control for the attacker. }
	\label{fig:practical3}
\end{figure}

Figure~\ref{fig:practical3} depicts this causal model; boxes in grey highlight
the Linderberg accountability pattern and the dashed box marks unobservable
variables. If we know what the attacker is doing, we can disentangle their
action from the pilot's actions and close the back-door path. The downside is
that we actually have to invest time and money into such an IDS and need to be
reasonably certain that it will log all relevant information about the pilot and
the attacker. 

However, if we can ensure that the RC cannot be affected by the attacker, we can
use the front-door criterion\footnote{See
\cite{fdexample} for a detailed example with linear models.} to estimate the
causal effect of the pilot on the UAV, even without installing an IDS. The
intuition behind this solution is that, because there is no back-door path from
\emph{Pilot} to \emph{RC}, we can estimate the effect of \emph{Pilot} on
\emph{RC}. Similarly, we can estimate the effect of \emph{RC} on \emph{UAV} and
calculate the effect of \emph{Pilot} on \emph{UAV}.  With such a causal
structure, we can do away with an IDS while still being able to ensure
accountability. The downside is that, beside the need to be sure that the causal
structure reflects the real world, the attacker must not be able to influence
the RC and the pilot must have no other way of affected the UAV than via the RC. 

Beside the back-door and front-door criterion, do-calculus offers us many
different options of analyzing and manipulating SCMs. If we express our system
as an SCM, similar to the examples given above, we can use do-calculus and tools
build for it \cite{tikka2018identifying} to either adapt the system to the
necessary structure or even find structures that make our system simpler.

\section{Discussion}

When looking at any of the figures in the paper, it is easy to suggest
additional variables, different connections, or new structures. Some readers
might look at the figures and argue that the model might be unrealistic or even
wrong. We do not claim that our examples are perfect or ``the one true
model''. On the contrary, we want to emphasize how easy these models are to
inspect, understand and criticize. The main advantage of using SCMs to express
accountability properties of systems is that they make definitions of
accountability easy to communicate and foster discussions. If we compare
Figure~\ref{fig:uber} to the story of the Uber accident in
Section~\ref{sec:formalizing}, we can see how a long and involved paragraph of
text can be expressed as an easy-to-discuss diagram. The exact details, such as
how Uber influences the driver, can be omitted in the graphical model and
expressed in the underlying SCM. For most use cases, first determining that
there is a causal influence from one node on another is enough.  Especially the
knowledge that some nodes do \emph{not} affect the outcome is useful to
establish accountability relationships between principals and agents. The core
advantage of using SCMs to express accountability properties, however, is that
they enable us to identify and communicate accountability patterns.   

Expressing accountability in SCMs is not without its problems. One obvious
problem is the creation of the models in the first place. There exists some work
on using other models of systems such as fault trees or Timed Failure
Propagation Graphs \cite{nfm}, as well as models of human interaction with
systems \cite{crest}, to generate the SCMs; however, so far these efforts are
semi-automatic and will require human input. Looking at the problems that
accountable algorithms have with fairness and bias, we think that this manual
input will not go away anytime soon. Without fully automatic solutions, we are
convinced that SCMs offer the best compromise between being understandable for
humans, while offering a rigid formal framework to analyze and reason over them.
The other major problem is defining the functions $\mathcal{F}$. While they can
take on any form, in many examples binary functions are used. The main reason
for this is that such models are computationally easier to analyze and also
straight-forward to explain. As in the model about Titus Manlius, binary models
can be interpreted as ``caused'' or ``do not cause''. If we now look at the Uber
example in Figure~\ref{fig:uber}, what could the function between Uber and the
driver look like? A binary function might just express that there is an
influence. This is good to know, but not necessarily enough. We might use a
probability and draw from observations that, for example\footnote{All these
examples are made up for illustrative purposes and are not true.},  95\% of the
drivers were instructed by Uber to do long hours and were thus not attentive.
But how does this help us, if we actually want to express the more subtle point
that Uber was rewarding miles driven and drivers were thus motivated to keep
driving when they were exhausted? When considering human interactions, it is not
hard to come up with examples that are hard to capture in neat closed-form
formulas. Here the big advantage of SCMs is that much analysis can be done on
models that do not specify  $\mathcal{F}$ explicitly. That is why we look at the
patterns of the graph and ignore the actual functions. If we ensure at the
design time of a system its the causal graph has a structure in which the causal
effects relevant for accountability are identifiable, we will usually be able to
find evidence to do so in a ex-post analysis. In other words, if we at least 
know that Uber has an effect on the driver, and build our system to log their
interaction, we will be able to find the exact effect in a post mortem analysis
where we have data on the system, as well as external observations available.

\section{Related Work}

Beside the works on accountability that we detail in
Section~\ref{sec:accountability}, accountability in computer science was
popularized by by Weitzner et al.~\cite{Weitzner:2008}, who proposed
``Information Accountability'' as a step beyond preventive data control
measures. They argue that it is impossible to keep data, especially personal
data, secret. Instead of trying to prevent data leaks, they suggest to build
accountable systems that make it easy to trace data leaks and use the legal
system to punish misbehavior. This work was continued by Feigenbaum~et al.~
\cite{feigenbaum2011towards,feigenbaum2012systematizing} with a focus on
security. Their definition of accountability is, for example, close to the
Linderberg pattern.  Feigenbaum~et~al. are, to the best of our knowledge, the
first to make a connection with causality, by pointing to the Halpern and Pearl
(HP) definition of causality \cite{halpern2005causes1,halpern2005causes2}. The
HP definition \cite{halpern2015} of causality is closely related to the
definition of causality we use here. However, while we here use a \emph{type
causal} definition, HP is used to define \emph{actual causality}. Type causality
is forward-looking and will often be used by scientific models. A classic
example of type causality is a statement like ``\emph{Cigarettes cause lung
cancer}'' \cite{halpern2016actual}. Actual causality, on the other hand, is used
to identify specific courses of events in the past. A classic example is the
statement ``\emph{the fact that David smoked like a chimney for 30 years caused
him to get cancer last year}'' \cite{halpern2016actual}. To build systems, we
need forward-looking, type causal models. Fortunately a good type causality
model will also be a good actual causality model and make it easy to analyze
events ex-post. For example, \cite{halpern2018towards} build on actual causality
to give definitions of blameworthiness and moral responsibility. An interesting
property of type causal models is transportability \cite{pearl2014external}.
It describes under what assumptions we can transfer causal knowledge
gained in experiments to new contexts where only observations are possible.   

There are various implementations of accountability mechanisms in the computer
science literature. Kacianka et al.~\cite{kacianka2017mapping} conducted a
systematic mapping study on implementation, but found that many papers use no
explicit definition of accountability. For the cloud domain, the A4Cloud
project~\cite{a4cloud2014} published an extensive report on accountability for
cloud providers. From other domains, K\"{u}sters et
al.~\cite{kusters2010accountability} give a mathematical definition of
accountability in the context of cryptography for e-voting systems, Datta et
al.~\cite{datta2016accountability} suggest to base accountability in CPS on
causal information flow analysis and  Baldoni et
al.~\cite{baldoni2018information} published an information model of
accountability as an Object-Role Model.  As in computer science finding the
causes of problems and blaming the responsible party is an old problem, many
techniques that are not specifically called accountability, yet solve similar
problems, were developed. For example, logging is used to understand system
behavior~\cite{amir2016correct}, runtime verification is used to ensure correct
system executing~\cite{leucker2009brief}, fault trees~\cite{vesely1981fault} are
a technique from the safety domain to express ways in which a system may fail,
fault localization~\cite{renieres2003fault} describes techniques to find faults
in source code and Forensics-by-Design~\cite{fbd2016} is a technique from the
security domain to build systems so that, if unexpected attacks happen, they can
still be detected. We are positive that all approaches that, in essence, ask
questions of accountability, can be express as an SCM and then analyzed for
patterns.  

A fairly new branch of accountability in computer science is algorithmic
accountability~\cite{diakopoulos2015algorithmic}. Here, the goal is to
understand the behavior of, often, machine-learning algorithms. Ananny and
Crawford~\cite{ananny2018seeing} have shown that mere transparency is not
sufficient for accountability.  Miller~\cite{miller2018explanation} gives a
thorough overview of the concept of an explanation and suggests that actual
causality is useful tool to create and give explanation. Doshi-Velez et
al.~\cite{doshi2017accountability} point to the importance of explanations of AI
systems in the legal context and highlight that explanations need to be causal.
Mittelstadt et al.~\cite{mittelstadt2016ethics} elaborate on the ethical
implications of AI systems and also point to the relevance of causality and
causal knowledge. Wachter et al.~\cite{wachter2017counterfactual} also suggest
using counterfactuals to explain the behavior of black box AI systems.
Mittelstadt et al. \cite{mittelstadt2019} recently struck a more careful note
and suggest that we need to be careful with models used to explain AI systems.
They especially highlight the utility of contrastive explanations for human
understanding. As SCM make it easy to explore counterfactual alternatives, we
expect that our approach of expressing accountability with SCMs can be adapted
to the problems of algorithmic accountability and explainable AI. Combining
these models with efficient actual causality reasoning~\cite{ibrahim2019}, work
on abstracting causal models~\cite{beckers2019abstracting} and efforts to
automatically derive SCMs for CPS~\cite{nfm}, will allow us to identify root
causes, give counterfactual explanations,  exchange models between different
domains of computer science and, most importantly, other fields of~science.

\section{Conclusions}
CPS will continue to be deployed in shared spaces with humans. This will lead to
questions of accountability of those systems for their actions and for those
systems by their manufactures, owners and operators. In the first part of this
paper, we looked at the concept of accountability from different disciplines and
concluded that causality is at the core of any accountability definition.
Following this, we turned to current research on causality, and especially SCMs,
and showed with examples that they are a useful tool to express accountability
definitions and that SCMs allow us to identify accountability patterns. Using
methods to check if a cause is identifiable, we explored how different system
designs can be adapted to fit specific patterns. Using an accountability pattern
will ensure that the relevant causal effects in our systems are identifiable. We
can then use this knowledge to decide what values to log and which events are
safe to ignore. 

While we have shown that SCMs are useful to communicate accountability
definitions in a system's design, actually developing these SCMs is still an
open problem. Additionally, reasoning about \emph{actual causes} in such models
suffers from a high computational complexity and can currently only be done
automatically for binary models. In that regard, SCMs have the advantage that
they are already used in fields such as statistics, data science, or
epidemiology and that we can use theoretical results and, especially, tools from
these field to analyze accountability models. We are convinced that the
adaptation of SCMs to express accountability definitions will improve the
communication with fields such as philosophy, sociology or psychology that have
a long history of analyzing and understanding the concept and that these
discussions will lead to the emergence of new accountability patterns, insights
and best~practices. 

\begin{acks}
This work was supported by the Deutsche Forschungsgemeinschaft (DFG) under
grant no. PR1266/3-1, Design Paradigms for Societal-Scale Cyber-Physical Systems.

\end{acks}

\bibliographystyle{ACM-Reference-Format}
\bibliography{bibliography.bib}

\end{document}